\documentclass[a4paper,10pt]{article}
\usepackage[utf8]{inputenc}
\usepackage{amssymb,amsmath}
\usepackage{authblk}
\usepackage[left=0.5in, right=0.5in, top=1in, bottom=1in]{geometry}
\usepackage{hyperref}

\title{A note on Fermi energy of Fermi gas trapped  under generic power law potential in $d$-dimension}
\author{Mir Mehedi Faruk\\
Department of Theoretical Physics, University of Dhaka, Dhaka-1000\\
\href{mailto:me@somewhere.com}{Email: muturza3.1416@gmail.com, mehedi.faruk.mir@cern.ch} 
 }

\begin{document}
\maketitle
 
 \begin{abstract}
Average energy per fermion in case of Fermi gas with any kinematic characteristic, trapped under most general power law potential in $d$ dimension has been calculated at zero temperature. In a previous paper (M. Acharyya, Eur. J Phys. 31 L89 (2010)) it was shown, in case of free  ideal Fermi gas as dimension increases average energy approaches to Fermi energy and in infinite dimension average energy becomes equal to Fermi energy at $T=0$. In this letter  it is shown that, for trapped system at  finite dimension the average energy depends on power law exponent, but as dimension tends to infinity average energy coincides with Fermi energy for any power law exponent. The result obtained in this manuscript is more general as we can describe free system as well as any trapped system with appropriate choice of power law exponent and true for any kinematic parameter.
 
 \end{abstract}
 .\\ \\ \\
The behaviour of ideal quantum gas are  studied in literature\cite{pathria, huang, ziff} widely where the
thermodynamic quantities as well as Fermi energy, average energy
per fermion are examined in great detail. In a real system, of course interaction between particles do exist. 
But taking it into account makes the problem difficult to
solve analytically. Thus it is well
approximated that the Bose gas of low density can be treated as in ideal bose gas. Neveretheless, to understand the effect
of interactions and to retain the essential physics, we approximately represent the real system by non interacting particles
in the presence of an external potential
The constrained role of external potential
for atomic gases do change the performence of gases.
Thus trapped atomic gases provide the opportunity to manipluate t
he quantum statistical effects\cite{turza,sala}.
An interesting conclusion was drawn in 
Ref.\cite{acharyya4} that in case of ideal free Fermi gas,
average energy approaches to Fermi energy with increament of dimension
and coincides with Fermi Energy when dimension tends to infinity.
So, it will intriguing to check this conclusion in case of trapped system. To do this we have took the most general
power law potential which is not essentially symmetric so that we can describe free system as well as any other trapped
system choosing suitable power law exponent. Moreover 
the invesigation is done with arbitrary kinematic parameter to obtain the more general result . As it turns out the result is true for any kinematic parameter.\\\\
Considering  ideal Fermi gas in a confining external potential in a $d$ dimensional space with energy spectrum, 
\begin{eqnarray}
E (p,x_i)= bp^l + \sum_{i=1} ^d c_i |\frac{x_i}{a_i}|^{n_i}
\end{eqnarray}
Where, the first part denotes the kinteic energy and the second part
stands for the trapping potential. Here, $b,$ $l,$ $a_i$, $c_i$, $n_i$  are all positive constants, $p$ is the momentum 
and $x_i$ is the  $i$ th component of coordinate of a particle. . Here, $c_i$, $a_i$ and $n_i$ determine the depth 
and confinement power of
the potential. For the free system all $n_i\longrightarrow \infty$.
$l$ is the kinematic parameter.
With $l=2$, $b=\frac{1}{2m}$ one can get the energy spectrum  of  the hamiltonian used in 
the literature \cite{pathria,huang,ziff}. And with $l=1$ and $b=c$, where $c$ is the velocity of light, we get case of ultrarelativistic Fermi system.\\\\
Density of states can be calculated from the following formula,
\begin{eqnarray}
 \rho(E)&=& \int \int \frac{d^d r d^d p}{(2 \pi \hslash)^d} \delta(E - E (p,r))
\end{eqnarray}\\
Now,  from Eq. (1), we obtain the density of states, \cite{turza}
\begin{eqnarray}
\rho(E)=C(m,V)E^{\chi-1} 
\end{eqnarray}
where, $C(m,V)$ is a constant depending on volume and mass\cite{turza} 
and $\chi=\frac{d}{l}+\sum_i ^d\frac{1}{n_i}$. As temperature $T\longrightarrow 0$,
the Fermi-Dirac distribution function reduces to,
\begin{eqnarray}
F(E)
   &=&\left\{
     \begin{array}{lr}
  1 &,E\leqslant E_F\\
  0 &,E > E_F
     \end{array}
   \right.
\end{eqnarray}
Now, we can easily calculate the average energy per fermion
\begin{eqnarray}
 \langle E \rangle 
 =\frac{\int_0 ^{E_F} \rho(E) E F(E)}{\int_0 ^{E_F} \rho(E) F(E)}= \frac{\chi}{\chi +1}E_F
 =\frac{\frac{d}{l}+\sum_i ^d\frac{1}{n_i}}{\frac{d}{l}+\sum_i ^d\frac{1}{n_i}+1}E_F
\end{eqnarray}
Above equation suggests average energy per fermion for trapped system depends on
dimension $d$ as well as power law exponent. So, in case of free system, all $n_i\longrightarrow \infty$ and 
the above expression reduces to (denoting average energy by $\langle E' \rangle$ for free system),
\begin{eqnarray}
\langle E' \rangle = \frac{d}{d+l}E_F
\end{eqnarray}
And in case of symmetric potential $n_1=n_2=..=n_i=..=n_d$, Eq. (5) becomes,
\begin{eqnarray}
\langle E \rangle  = \frac{\frac{d}{l}+\frac{d}{n}}{\frac{d}{l}+\frac{d}{n}+1} E_F
= \frac{\frac{1}{l}+\frac{1}{n}}{\frac{1}{l}+\frac{1}{n}+\frac{1}{d}} E_F
\end{eqnarray}
In case of harmonic potential ($n_1=n_2=..=n_i=..=n_d=2$) average  energy  stands from Eq. (7) (choosing $l=2$),
\begin{eqnarray}
\langle E \rangle &=&\left\{
     \begin{array}{lr}
  \frac{1}{2}E_F &,d=1\\\\
  \frac{2}{3}E_F &,d=2\\\\
  \frac{3}{4}E_F &,d=3
     \end{array}
   \right.
\end{eqnarray}
In case of free Fermi system  it seen from Eq. (6) average energy is (choosing $l=2$),
\begin{eqnarray}
\langle E' \rangle &=&\left\{
     \begin{array}{lr}
  \frac{1}{3}E_F &,d=1\\\\
  \frac{1}{2}E_F &,d=2\\\\
  \frac{3}{5}E_F &,d=3
     \end{array}
   \right.
\end{eqnarray}
It is noteworthy that,
for both free and trapped system average energy per fermion tends to approach
Fermi energy with increament of dimension $d$.
Another noteworthy observation, there is a shift in average energy due to trapping potential. Also
at any specific dimension $d$, average energy  of trapped system 
gets more close to $E_F$ compared to average energy of free system. For instance at $d=3$, in case of free system
average energy per fermion is 
$60\%$ of Fermi energy, whereas for trapped system average energy per fermion 
is $75\%$ of Fermi energy. One can obtain the exact expression of shift of average energy 
for arbitrary dimension due to trapping potential.
Rewriting the expression Eq. $(7)$ for free system with $l=2$,
\begin{eqnarray}
 \langle E' \rangle = \frac{d}{d+2}E_F
\end{eqnarray}
In case of harmonic potential the average energy with $l=2$,
\begin{eqnarray}
 \langle E \rangle = \frac{d}{d+1}E_F
\end{eqnarray}
So, the shift of average energy due to external harmonic potential,
\begin{eqnarray}
\langle \Delta E \rangle = \langle E \rangle -\langle E' \rangle  =\frac{d}{(d+1)(d+2)}E_F
\end{eqnarray}
Obviously, one can have such relation for any $n_i$. Now, 
very interesting conclusion was drawn in Ref. \cite{acharyya4} that in free system 
as $d$ tends to infinity, average energy tends to Fermi energy,
 suggestsing that at infinite dimension each fermion has its energy
equal to Fermi energy.  Eq. (12) suggests as $d$
tends to infinity, shift of average energy 
becomes zero. So from this, one can certainly conclude for the systems trapped by harmonic potential,
average energy approches Fermi energy as $d$ tends to infinity which is also suggested by Eq. (11).
As it turns out, it is not only property of free or  trapped system with harmonic potential, but 
any system trapped with generic power law potential (not essentially symmetric).
Now, from Eq. (5) it is also seen,
as $d\longrightarrow \infty$, $\langle E \rangle \longrightarrow E_F$. So, the following theorem can be obtained.\\ 

\textbf{Theorem}: {\em For ideal Fermi gas with any kinematic characteristic is trapped  under generic power 
law potential $U(r)=\sum_{i=1} ^d c_i |\frac{x_i}{a_i}|^{n_i}$, as $d\rightarrow \infty$,
$\langle E \rangle = E_F$}. \\\\
As this generic potential can describe any other system with appropriate choice of $n_i$, this is a general statement with any 
kinematic characteristic parameter $l$. As $n_i\rightarrow \infty$, it coincides with the conclusion of Ref. \cite{acharyya4}. It means the conclusion of Ref. \cite{acharyya4} is a special case of the above theorem.
So, with the most general trapping potential we can find all fermions move with Fermi momentum $p_F=\sqrt{(2mE_F)}$ at $T=0$.
This suggests if we try to accomodate finite number of fermions in dimension $d$, the number of fermions lying in the surface 
increases as we increase dimensionalilty. And for a fixed number of particles all
fermions  do stay on Fermi hypersphere if the space dimensionalilty is infinity. 
But of course it does not violate
Pauli exclusion principle as Eq. (3) suggests number of energy states increases as dimensionalilty increases.
It concludes the fact that the fermions of trapped Fermi gas with any kinematic parameter
can easily be accomodated in Fermi energy when $d=\infty$ at zero temperature. It would be interesting to extend the 
study at $T\neq 0$. This theorem and its consequences are not yet discussed in the standard textbooks of quantum statistics.
I hope this will be helpful for students as well as researchers.
\section{Acknowledgement}
I would like to thank Fatema Farjana and Mishkta Al Alvi to point out the typographic mistakes.

  \end{document}